\documentclass[12pt]{iopart} 
\usepackage{graphicx,epsfig}
\begin{document}

%{\large \bf \underline{Characterization of the response of superheated}}
%\vspace{0.3cm}
%{\large \bf \underline{droplet (Bubble) detectors}}
%\vspace{0.5cm}

\title{Characterization of the Response of Superheated Droplet (Bubble) Detectors}

\author{M.Barnab\'e-Heider, M.Di Marco, P.Doane, M-H.Genest, R.Gornea, C.Leroy\footnote{\uppercase{C}orresponding author: leroy@lps.umontreal.ca}, L.Lessard, J.P.Martin, U.Wichoski, V.Zacek}

\address{Groupe de Physique des Particules,D\'epartement de Physique,Universit\'e de Montr\'eal,C.P.6128,Succ.Centre-Ville,Montr\'eal (Qu\'ebec) H3C 3J7, Canada}

\author{A.J.Noble}

\address{Department of Physics, Queens University\\
Kingston (Ontario) K7L 3NG, Canada }

\author{E.Behnke, J.Behnke, W.Feighery, I.Levine, C.Mathusi, J.Neurenberg, R.Nymberg}

\address{Department of Physics and Astronomy, Indiana University South Bend \\
South Bend, Indiana, 46634, USA }

\author{S.N.Shore}

\address{Dipartimento di Fisica "Enrico Fermi",Universtit\`a di Pisa,Pisa,I-56127,Italia}

\author{R.Noulty, S.Kanagalingam}

\address{Bubbles Technology Industries, Chalk River (Ontario) K0J 1J0, Canada }

\author{\bf \underline{PICASSO COLLABORATION}}

\abstract{
The PICASSO project is a cold dark matter (CDM) search experiment 
relying on the superheated droplet technique. The detectors use 
superheated freon liquid droplets (active material) dispersed 
and trapped in a polymerized gel. This detection technique is based
on the phase transition of superheated droplets at room or
moderate temperatures. The phase transitions are induced by nuclear
recoils when undergoing interactions with particles, including CDM
candidates such as the neutralinos
predicted by supersymmetric models. The suitability of the technique for this purpose
has been demonstrated by R\&D studies performed
over several years on detectors of various composition and volume. Simulations performed to understand the
detector response to neutrons and alpha particles are presented along
with corresponding data obtained at the Montreal Laboratory.}

\section{Introduction}

Superheated droplet detectors, here referred to as "bubble detectors", use superheated liquid droplets (active medium) 
dispersed and suspended in a polymerized gel \cite{apfel},\cite{bti}. Presently,
these droplet detectors consist of an emulsion of metastable
superheated droplets of liquids (such as C$_3$F$_8$, C$_4$F$_{10}$, CF$_{3}$Br, CCl$_2$F$_2$) at a temperature higher than their boiling point,
dispersed in an aqueous solution, subsequently polymerized after dissolution 
of an appropriate concentration
of a heavy salt (e.g., CsCl) in water to equalize
densities of droplets and solution. 
By applying an adequate pressure, the boiling temperature can be raised
allowing the emulsion to be kept in a liquid state.
Under this external pressure, the detectors are 
insensitive to radiation. 

By removing the external pressure, 
the liquid becomes superheated and sensitive to radiation. Bubble
formation occurs through liquid-to-vapour phase transitions, 
triggered by the energy deposited by radiation. This can be the result
of nuclear recoils
(through interactions with neutrons or other particles), or through
direct deposition (gamma, beta, alpha particles). Bubble detectors are
threshold detectors as they need to achieve a minimal energy 
deposition to induce a phase transition. Their sensitivity to various 
types of radiation strongly depends on the operating temperature and pressure. 
The liquid-to-vapour transition is explosive in nature and is accompanied by an 
acoustic shock wave which can be detected 
with piezoelectric sensors. The signal measurement of the detectors is described in \cite{razvan}
These detectors are re-usable by re-compressing the bubbles back to droplets. 

Bubble detectors of small volume, typically 10 m$l$, representing 0.1-0.2 g of active
material have been used for many years in applications such as portable neutron 
dosimeters for personal dosimetry
or the measurement of the radiation fields near particle
accelerators or reactors.

\begin{figure}[htb]
\begin{center}
\epsfxsize=5cm\epsffile{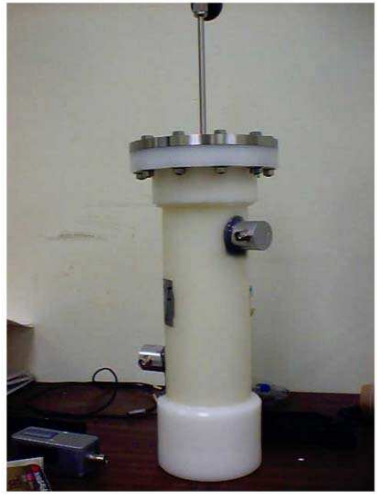}
\end{center}
\caption{A large-volume droplet detector module (1-litre volume) 
equipped with piezoelectric sensors glued on the surface for signal detection.
The container can be pressurized up to 10 bars. Typical C$_x$F$_y$ gas loading presently achieved for this type of detector is in 
the 5-10 g/litre range. 
}
\label{f:detector}
\end{figure}

Over the last several years, Bubble Technology Industries (BTI) and the
University of Montreal have been collaborating in the making of bubble
detectors using detector formulations that are more appropriate for the
application of the Picasso group.  To distinguish these detectors from the
conventional commercial bubble detectors, we have named these detectors
"Special Bubble Detectors (SBD)". More recently, the PICASSO group has developed
detectors of larger volume of the type shown in Fig.(\ref{f:detector}) with the aim to
perform a direct measurement of neutralinos predicted by minimal 
supersymmetric models of cold dark matter (CDM) particles. The very low interaction
cross section between CDM and the nuclei of the detector's active medium requires the use
of very massive detectors to achieve a sensitivity level allowing the detection 
of CDM particles in the 
galactic environment\cite{astro}.

\section{The SBD's and their operation}
 
The response of the SBD's to incoming particles or radiation 
is determined by thermodynamic properties of the active gas, such as operating temperature and pressure. The detector operation 
can be understood in the framework of the theory of Seitz\cite{seitz} which 
describes bubble formation as being triggered by the heat spike produced by
the energy deposited when a particle traverses a depth of superheated medium.

The boiling temperature, T$_b$, increases with pressure. The superheated condition
is achieved when the temperature of the liquid phase is higher than its boiling
temperature.
Active liquids at room temperature, but subjected to high enough pressures,
are not superheated. However, if the pressure is lowered to atmospheric
value, the liquids become superheated in the absence of heterogeneous
nucleation. In SBD's, over the temperature range
from boiling to critical temperatures 
(defined as the temperature at which the surface tension becomes zero), there
is a pressure threshold above which the droplets will be thermodynamically stable and the 
detector will not be sensitive. If the pressure is 
released or considerably decreased, droplets become superheated and the 
detector becomes sensitive.
 
The droplet should normally make
a transition from the liquid (high potential energy) to the gaseous
phase (lower potential energy). However, undisturbed, the droplet is in a metastable state since it must overcome a potential barrier to make the transition 
from the liquid to the gas phase. This can be done if the droplet receives 
an extra amount of energy such as the heat due to the energy deposited by incoming particles. The potential barrier is given by Gibbs' equation
\begin{equation}
E_c~=~{{16\pi}\over{3}}{{\sigma(T)^3}\over{(p_i~-~p_0)^2}}  ,
\label{eq-gibbs}
\end{equation}
where $p_0$ and $p_i$, the externally applied pressure and the vapour pressure in the bubble, respectively, are temperature dependent. The difference between
these two pressures represents the degree of superheat.
The surface tension of the liquid-vapour interface at temperature $T$ is given 
by $\sigma(T)$~=~$\sigma_0 (T_c~-~T)/(T_c~-~T_0)$ where $T_c$ is the critical 
temperature of the gas, $\sigma_0$ is the surface tension at a reference 
temperature $T_0$, usually the boiling temperature $T_b$. For a
combination of two gases, $T_b$ and $T_c$ can be adjusted, depending
on the mixture ratio. 
For instance $T_b$ = -19.2$^{o}$C, $T_c$ = 92.6$^{o}$C for a SBD-100
detector,(loaded with a mixture of fluorocarbons); whereas 
$T_b$ = 1.7 $^{o}$C, $T_c$ = 113.0 $^{o}$C for SBD-1000 detectors (loaded with
100$\%$ $C_4F_{10}$) \cite{nadim}. 
The response of various types of SBD's can be compared by using the
reduced superheat variable, s, defined as $s= \left(T-T_{b}\right)/\left(T_{c}-T_{b}\right)$.

Bubble formation will occur when a minimum deposited energy,
$E_{Rth}$, 
exceeds the threshold value $E_c$ within a distance $l_c$~=~$a$ $R_c$, where the critical radius $R_c$ is given by:
\begin{equation}
R_c~=~{{2\sigma(T)}\over{(p_i~-~p_0)}}  .
\label{Eq-Rc}
\end{equation}
A value $a$ $\approx$ 2 is suggested in Ref.\cite{apfel2}, but higher values, up to 13, are given in Ref.\cite{bell}.
If dE/dx is the mean energy deposited per unit distance
(this energy is a function of the energy of the nuclei recoiling after their
collision with the incident particle), the energy
deposited along $l_c$  
is $E_{dep}$~=~dE/dx $\times$ $l_c$. 
The condition to trigger a liquid-to-vapour
transition is $E_{dep}~\geq~E_{Rth}$.
Since it is not the total deposited energy that will trigger a liquid-to-vapour transition, but the fraction of this energy transformed into heat, the actual minimum or threshold energy, $E_{Rth}$, for recoil detection is related to $E_c$ by
an efficiency factor, $\eta$~=~$E_c/E_{Rth}$ ($2<~\eta~<6\%$)\cite{apfel,harper}.
%an efficiency factor $b$ has to be accounted for and the 
%minimum energy, $E_{min}$, for recoil detection is
%\begin{equation}
%E_{min}~=~b~E_c
%\label{Eq-emin}
%\end{equation}
The threshold value being dependent on the operating temperature and pressure, 
the detector can be set into a regime where it responds mainly to nuclear
recoils and discrimination can be achieved against background
radiations from minimum ionizing particles and gamma rays. 

\section{Alpha-particle response measurement and simulation}

The heavy salt and other ingredients, mixed in the gel at the present stage of detector
production, contains contaminants which are $\alpha$-emitters, such as U/Th and daughters. 
The $\alpha$ background is one of the main backgrounds at normal
temperatures of SBD operation since other backgrounds contribute 
to the detector signal
predominantly at higher temperatures.(Neutrons, discussed in the next section, will
be greatly shielded against by operating the detectors deep underground). Therefore, the response of SBD to $\alpha$-particles has  
to be investigated carefully.

\begin{figure}[htb]
\begin{center}
\epsfxsize=9cm\epsffile{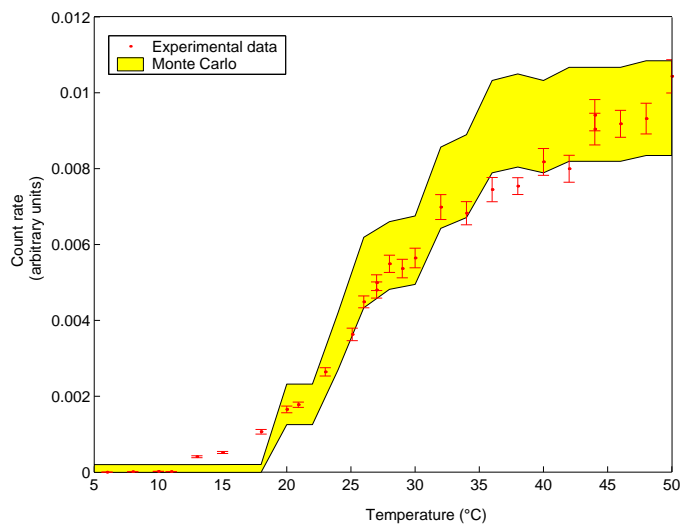}
\end{center}
\caption{Detector response (count rate) of the 1$l$ SBD-1000 spiked with 20 Bqs of $^{241}$Am as a function of temperature. A critical length of L=18R$_c$ is necessary to fit the data.}
\label{f:XXS} 
\end{figure}

The alpha response was studied using a 1$l$ SBD-1000 fabricated at BTI
(Bubble Technology Industry). In the fabrication process, 27.8 m$l$ of
an americium solution (AmCl$_{3}$ in 0,5M HCl) of known activity
(0,7215 Bq/m$l$) were added to the gel 
solution to create a detector spiked with 20 Bqs of $^{241}$Am. Then,
the detector liquid was added, followed by mixing and polymerisation
procedures. Using this detector, the alpha detection efficiency was
measured at temperatures ranging from 6$^{o}$C to 50$^{o}$C (see
Fig.~(\ref{f:XXS})).  The alpha efficiency is defined as $\epsilon =
{{\tau}\over{A}}$, where $\tau$ is the count rate in bubbles per
second and $A$ is the activity of the source in Bqs. The efficiency is
temperature dependent.
% and seems to level off at 1$\%$ around 45$^{o}$C.  

 In the temperature range studied the detector is essentially  insensitive to $\gamma$ and $\beta$ radiation and the neutron background contribution is smaller than the errors bars. 

\begin{figure}[htb]
\begin{center}
\epsfxsize=7cm\epsffile{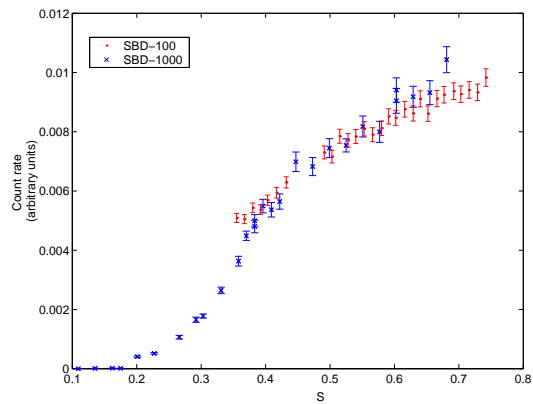}
\end{center}
\caption{Detector response (count rate) of the 1$l$ SBD-1000 and SBD-100, spiked with 20 Bqs of $^{241}$Am, as a function of reduced superheat.}
\label{f:XXC} 
\end{figure}

A similar alpha measurement has been done with a SBD-100, but the
threshold temperature could not be reached. The higher operating
temperature of the SBD-1000 allows us to explore the threshold region
for the first time.
Note that the SBD-1000 and SBD-100 detectors have the same response
curve when the reduced superheat is used to describe the temperature dependence (see Fig.~(\ref{f:XXC})).

The alpha response was simulated, using Geant 4.5.2\cite{geant}. 
The geometry of the virtual detectors was approximated to an 8 m$l$
cylinder (the first generation of detectors were built in vials) 
made out of polycarbonate and filled with a gel loaded with CsCl. 
This approximation was valid even for the 1$l$ detector. 
The correction of the response due to a different surface to volume
ratio was calculated to be negligible 
due to the very short range of the alpha particles. 
The density of the superheated droplets was varied as a function of
temperature and the droplets were dispersed 
randomly in the gel. For all simulations the experimentally known
distribution of droplet diameters was used\cite{nadim}. 
The loading was assumed to be 0.7$\%$, a choice validated by a comparison between the neutron efficiency of the simulated detector and another 1$l$ detector for which the droplet distribution and loading were directly measured. Alpha particles were generated randomly in the gel, with an energy spectrum corresponding to the $^{241}$Am decay.  

The tracking of each particle or nucleus was done step by step and
when any track entered a droplet, 
the output file was updated by adding a code identifying the particle or nucleus involved, the energy it deposited and 
the corresponding deposition length. The ionization of low energy
nuclei was taken into account, 
using the ICRU$\_$R49\cite{icru} nuclear stopping model and the
SRIM2000p\cite{trim} 
electronic stopping model. An external analysis code was then used to
calculate the maximal energy each particle or 
nucleus deposited inside the critical length. Finally, this energy
value was used as input to a probability distribution function, described below (see Eq.\ref{prob}), to decide whether a vaporization would occur or not. Different requirements for minimal energy deposition, critical length and probability function have been studied in this analysis.

The minimal energy deposition needed to trigger a vaporization is known from the 
experimental threshold curves obtained with neutrons (see Fig.(~\ref{f:temp-press})), assuming a head-on collision between the neutron and a nucleus inside the droplet. The energy deposition efficiency $\eta$, described in Section 2, has been calculated with the method described in Ref.\cite{apfel2} to verify the consistency of the results with published values. The value obtained at 20$^{o}$C, assuming a recoiling carbon, is $\eta$=4$\%$, a result consistent with the 2-6$\%$ range generally accepted. The threshold energy considered in this calculation is the energy that the recoiling nucleus can deposit within one critical diameter (2$R_{c}$), a length smaller than the range of the recoiling nucleus considered. Therefore, this method of calculation assumes that the energy deposition necessary is less than the recoiling energy of the nucleus.  To maintain the coherence between neutron and alpha measurements, the simulations strongly suggest that all the recoiling energy is needed. The critical length for the necessary energy to be deposited will hence be greater than the range of the nucleus considered.

The probability, P($E_{dep}$, $E_{Rth}$), that a recoil nucleus at an energy near threshold will generate an explosive droplet-bubble transition is described by\cite{nadim}
\begin{equation} 
\label{prob}
%P = 1-\exp\left[{- b({E_{dep}-E_{Rth}})/{E_{Rth}}}\right]  ,
P = 1-\exp\left[\frac{- b(E_{dep}-E_{Rth})}{E_{Rth}}\right]  ,
\end{equation}
where $b$ is a free parameter to be fit.

Monte Carlo studies of the alpha response indicate that the
experimental efficiency is too high for the vaporization to be caused 
only by elastic collisions between alpha particles and nuclei in the
droplets. This leads to the suggestion that the phase transition 
is triggered by the alpha particles' ionization loss in the
droplets. Since the americium solution used in the spiked detector 
fabrication is hydrophilic and since the freon droplets are
hydrophobic, 
we can assume that the americium doesn't diffuse in the
droplets. Furthermore, the experimental efficiency is low enough to
consider no surfactant effect, as described in
ref.\cite{pan}. Under those assumptions, the contribution of the
recoiling short-range 
nucleus can be neglected ($^{237}$Np in the case of $^{241}$Am decays). 

Under the assumption that the recoiling nucleus triggering
vaporization at neutron threshold is fluorine, 
the dE/dx required to trigger a phase transition is too high to
explain the efficiencies seen in the alpha case. 
This is not completely understood, however it suggests that the
minimal energy deposited at neutron threshold must be defined by the
carbon recoil, as assumed in calculating the efficiency $\eta$. 
Taking the probability function (Eq.\ref{prob}) and this minimal energy
requirement, 
the critical length as a function of temperature and the value of $b$
were deduced from the fit of the data (see Fig.~(\ref{f:XXS})). 
The value of the critical length obtained is L=18$R_{c}$ and $b$=1. A
68.27$\%$ C.L interval for the Poisson 
signal was calculated for a simulated number of counts lower than
21. The error becomes $\sqrt{N}$ for a larger number of events. 

One can do further tests to check these assumptions by simulating the
neutron response to see if the result 
is consistent with the alpha case, using the same energy deposition and critical length requirements.

\section{Neutron response measurement and simulation}
SBD's have a high efficiency for detecting neutrons while being
insensitive to minimum ionizing particles and to nearly all sources of background 
when operated at the temperature and pressure region of interest for neutralino recoil detection\cite{nadim}. 

From purely kinematical considerations, nuclear recoil thresholds in SBD's
can be obtained in the same range for neutrons of low energy 
(e.g., from 10 keV up to a few MeV)and massive neutralinos (60
GeV/$c^2$ up to 1 TeV) with no sensitivity to minimum ionizing
particles and gamma-radiation. Therefore, the SBD response to neutrons has to be studied carefully. 

\begin{figure}[htb]
\begin{center}
\epsfxsize=7cm\epsffile{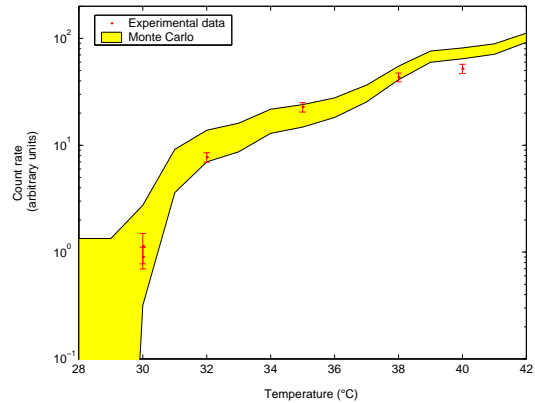}
\end{center}
\caption{SBD-1000 response (count rate) to 200 keV neutrons as a function of 
temperature. The volume of the detector is 8 m$l$. The simulated response gives a loading of 0.61$\pm$0.06$\%$}
\label{f:XXY} 
\end{figure}

\begin{figure}[htb]
\begin{center}
\epsfxsize=7cm\epsffile{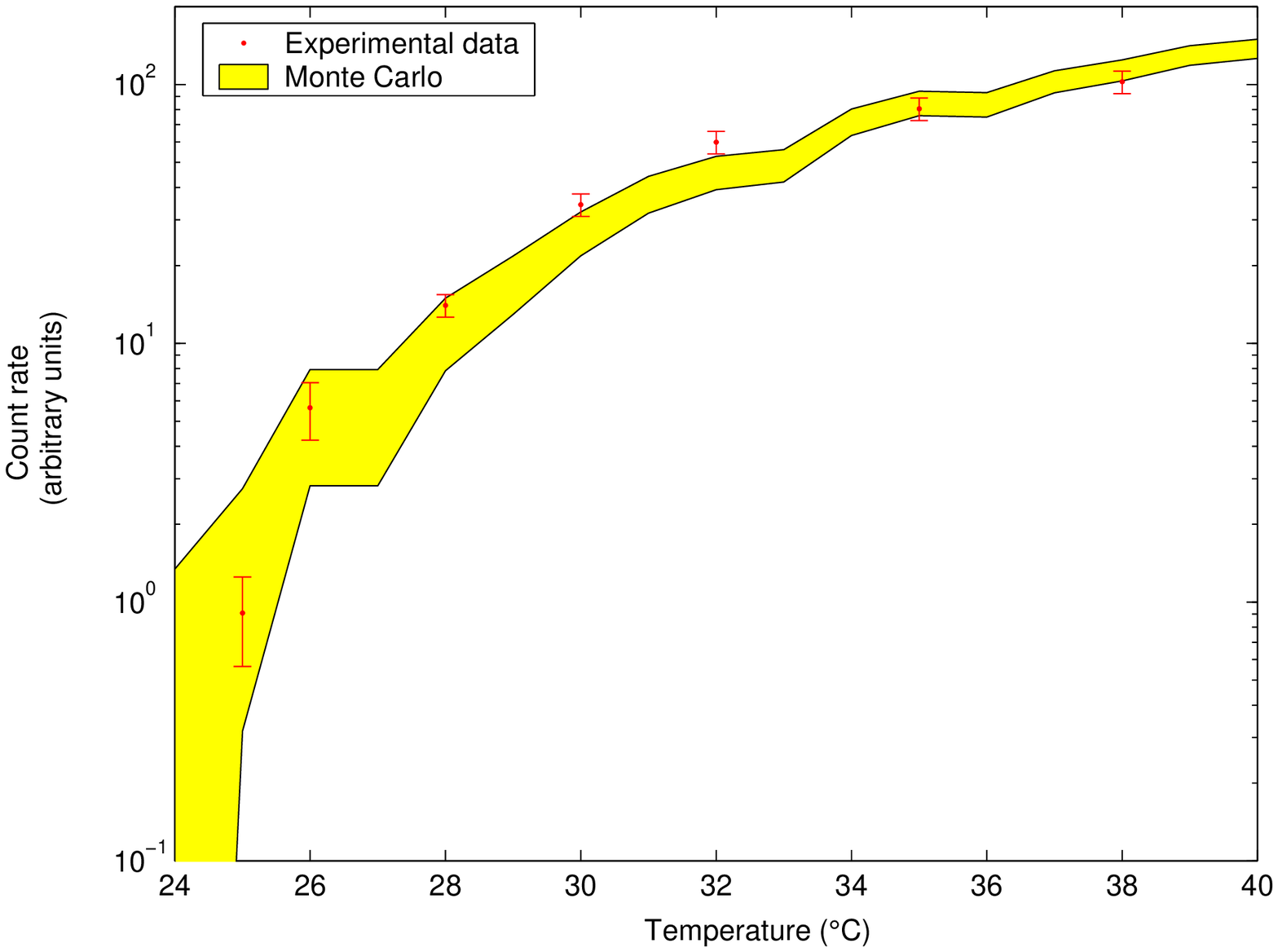}
\end{center}
\caption{SBD-1000 response (count rate) to 400 keV neutrons as a function of
temperature. The volume of the detector is 8 m$l$. The simulated response gives a loading of 0.69$\pm$0.03$\%$}
\label{f:XXZ} 
\end{figure}

\begin{figure}[htb]
\begin{center}
\epsfxsize=8cm\epsffile{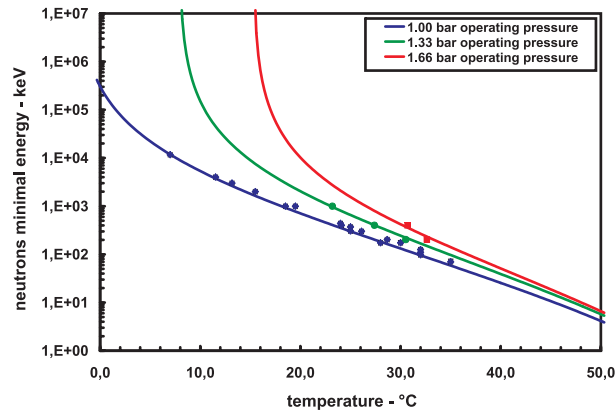}
\end{center}
\caption{Neutron threshold energy ($E_{Rth}$) as a function of temperature for various
operating pressures\cite{como2001}.}
\label{f:temp-press}
\end{figure}

The SBD response to mono-energetic neutrons has been measured as a function
of the temperature. These mono-energetic neutrons are obtained from
$^7Li(p,n)^7Be$ reactions at the Tandem van der Graaff facility of the Universit\'e de Montr\'eal. The detector response (count rate) to mono-energetic
neutrons of 200 and 400 keV as a function of operating temperature are shown 
in Figs.~(\ref{f:XXY}) and (\ref{f:XXZ}), respectively, for a detector 
of 8 m$l$ volume loaded with 100\% $C_4F_{10}$ droplets. 
From such curves, one can extract the threshold temperature, $T_{th}$, for a given neutron energy by extrapolating the curves to the lowest point. Then, it is possible to represent the neutron threshold energy as a function of temperature (Fig.(~\ref{f:temp-press})).

The response of an 8 m$l$ SBD-1000 detector to mono-energetic neutron beams was also simulated, also using Geant4 and external analysis codes as in the alpha case. The loading was set to be 0.7$\%$ and, after the analysis, the response was normalized by a multiplicative factor to fit the experimental data. This factor was interpreted as the loading correction from the initially assumed 0.7$\%$ value. 
The energy variation in the proton beam used to create neutrons causes
the neutron energy to fluctuate around the 
beam energy mean value. Consequently, the neutrons generated in the simulation were given an initial energy with Gaussian fluctuations, e.g. $\sigma$=5 keV for 200 keV neutrons. Neutron elastic and inelastic scattering were considered for all types of nuclei using the ENDF/B-VI data library\cite{endf}, in all detector parts.
As can be seen in Figs.~(\ref{f:XXY}) and~(\ref{f:XXZ}), the simulated response fits well the experimental data for E = 200 and 400 keV, validating the critical length and minimal energy requirements used to fit the alpha response. Fitting the neutron experimental data also allows the determination of the detector's loading. The values of 0.61$\pm$0.06$\%$ and 0.69$\pm$0.03$\%$ obtained at 200 and 400 keV respectively are consistent with each other.

\section{Expected Neutralino Response}

Due to its $^{19}$F content, PICASSO's detectors are especially suitable to search for spin-dependent neutralino interactions. $^{19}$F is a spin-${1\over2}$$^{+}$ isotope and has a very favorable spin dependent cross-section\cite{ellis}. Knowing the $^{19}$F recoil spectra expected for interactions with neutralinos in our galactic halo and knowing the detector threshold for fluorine recoils at a given operating temperature as a function of temperature, we can determine the neutralino induced count rate for a given interaction cross section and neutralino mass at different temperatures. Thus, although the superheated droplet detector is a threshold device, spectral information can be obtained by ramping the temperature. 

\begin{figure}[htb]
\begin{center}
\epsfxsize=7cm\epsffile{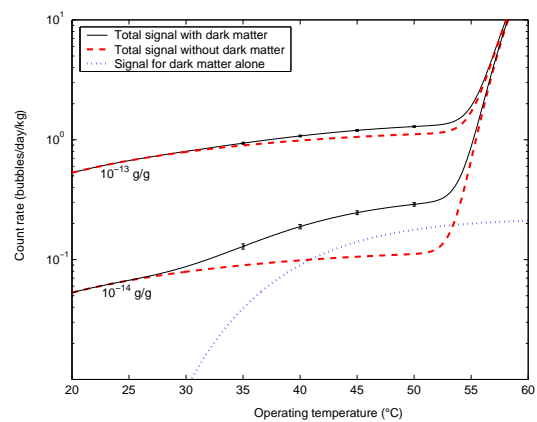}
\end{center}
\caption{Expected signal for a neutralino of 100 GeV/c$^{2}$ and a cross section of 0.004 picobarn.}
\label{f:XNEUT}
\end{figure}

The expected signal for a neutralino of 100 GeV/c$^{2}$ and a cross
section of 0.004 pb is shown in Fig.~(\ref{f:XNEUT}) for  a detector
of 100 kg of 
active mass and 30 days running time. For these calculations, we have
assumed 
canonical values of 0.3 GeV/cm$^3$ for the local neutralino density in
the solar system and a mean velocity of 235 km/s for the 
Maxwellian velocity distribution. In the recoil signal temperature
range (from 20$^{o}$C to 50$^{o}$C), the main 
contribution to the background is due to the alpha particles resulting
from U/Th contaminations in the detector itself. The contribution from
neutrons will be low for installations deep underground and with
adequate shielding.
The sensitivity to minimum ionizing particles and gamma-rays is
expected to occur above 50$^{o}$C. 
As can be seen in Fig.~(\ref{f:XNEUT}), the signal becomes clearly visible at a background contamination level of 10$^{-14}$g/g U/Th. The knowledge of the precise shape of the alpha response allows us to extract either the signal or an upper bound to the neutralino cross-section. Presently the contamination level in the detectors is at the level of approximately 10$^{-10}$g/g. Work is in progress to reduce this contribution substantially in the near future.

\section{Conclusions}
   Simulations performed to understand the response of superheated
droplet detectors to neutrons and alpha particles indicate that all
the data can be well described with a unique, consistent set of
variables which parameterize the underlying model of recoil energy
threshold and energy deposition (Seitz theory). The simulations of the
neutron data are able to trace the response of mono-energetic neutrons
as a function of temperature over several orders of magnitude in count
rate. The experimental alpha response is equally well described and
the simulations show that it is the alpha particles which mainly
trigger droplet formation by their specific energy loss along the
track, rather than the recoiling $^{237}$Np nucleus. 
The detection efficiency at higher
temperatures is consistent with the assumption that the alpha emitters are distributed uniformly in the gel surrounding the droplets.
Further tests and simulations with U and Th will be carried out
soon. These cases are complicated by the fact that these chains need
not be in secular equilibrium, and the radon daughters may diffuse
through the detector.   

   On the experimental side, we found a consistent description of the
alpha response 
for a series of alpha spiked detectors with different active detector
liquids and therefore different operating temperatures, if represented in terms of reduced superheat.  The range of alpha sensitivity coincides with the range of temperatures where droplet detectors are sensitive for neutralino induced recoils, but the shape of the response curves are different. The precise knowledge of the alpha response is therefore important in order to increase the sensitivity for neutralino detection in the presence of alpha emitting contaminants.   
      
   In the meantime a new phase of the project has started with the production of larger scale 1$l$ modules and their installation and operation at the location of the SNO detector in the Creighton mine at Sudbury, Ontario. Ten detector modules have been installed so far and operated over a period of several months, with their data analysis being in progress. 

\end{indented}
\end{document}